# Evidence of the impurity spin coupling with quantum paraelectric fluctuations in CaTi$_{1-x}$Ru$_x$O$_3$

Ivica M. Bradarić[1]* and Feodor V. Kusmartsev[2]

[1] Laboratory for Theoretical and Condensed Matter Physics, "Vinča" Institute of Nuclear Sciences, University of Belgrade, P.O. Box 522, 11001 Belgrade, Serbia.

[2] Department of Physics, Loughborough University, Loughborough, LE11 3TU, UK.

*Correspondence to: bradaric@vin.bg.ac.rs.

**Quantum paraelectrics are materials in which a long-range ferroelectric/antiferroelectric order is suppressed by quantum fluctuations, i.e. zero-point motion of the lattice prevents condensation of the soft polar phonon mode even at T = 0 K. The most prominent quantum paraelectric materials are SrTiO$_3$, KTaO$_3$, and CaTiO$_3$. Here we focus on peculiar properties of the pseudo-cubic perovskite CaTi$_{1-x}$Ru$_x$O$_3$ system. Namely, as soon as any concentration of either Ru or Ti is introduced into the pure compounds, a concentration-independent ferromagnetic-like transition occurs at low temperatures. We present the experimental evidence of the spin-polarized ground state of CaTi$_{1-x}$Ru$_x$O$_3$ induced by coupling of magnetic moments of Ru impurities with quantum paraelectric fluctuations in the host compound CaTiO$_3$.**

The prospect of developing the new quantum technologies, which promise qualitatively new level of ultrafast computing, akin to human brain, provided enormous boost to already exciting research in quantum materials. Fluctuations in quantum materials, which arise due to Heisenberg's uncertainty principle, drive phase transitions at zero temperature - quantum phase transitions. By adjusting appropriate quantum tuning parameter for the system at hand (pressure, chemical substitutions, magnetic field, electrical field, etc.), quantum fluctuations grow to the point (quantum critical point - QCP) where the long range order is prevented even at zero temperature (*1*). Investigations of quantum phase transitions and quantum criticality have been recently expanded to include dielectric materials, i.e. quantum paraelectrics (QPE) SrTiO$_3$ and KTaO$_3$ (*2-4*), see Fig.1.A. These materials belong to the displacive type ferroelectrics, but quantum fluctuations of electric dipoles prevent classical ordering at any non-zero temperature (*5*). Because the exotic phases of matter frequently appear close to QCP, discovery of superconductivity in SrTiO$_{3-\delta}$ (*6*), motivated search for superconductivity emerging in the vicinity of ferroelectric instability in many different systems (*7-11*). To this end, recent theoretical and experimental investigations indeed show decisive role of QPE fluctuations as driving mechanism of superconductivity in SrTiO$_{3-\delta}$ (*12, 13*). Consistent with these findings, optically generated carriers in SrTiO$_3$ show strong coupling to QPE fluctuations (*14*). On the other hand, coupling of QPE fluctuations with magnetic moments has been less explored. However, recent experiments on geometrically frustrated organic system κ-H$_3$(Cat-EDT-TTF)$_2$, present existence of quantum electrical dipole liquid, which, via coupling of localized spins with QPE fluctuations, leads to exotic quantum spin liquid state (*15*).

Here we show that even the slightest $Ru^{4+}$ substitution in $CaTiO_3$ prompts "ferromagnetic-like" transition below $T^* \approx 35\ K$ concurrently with the onset of QPE fluctuations, thus evidencing coupling of the impurity spin with QPE fluctuations in the host compound. Furthermore, these results imply the existence of the phase transition from paraelectric to quantum paraelectric state (see Fig. 1.A).

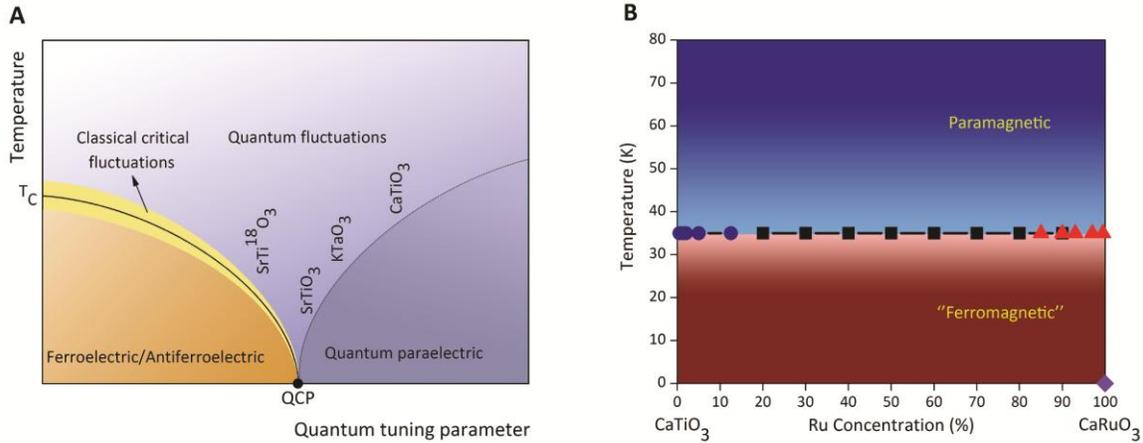

**Fig. 1. Phase diagrams.** (**A**) Schematic view of the quantum phase diagram for displacive ferroelectrics. (**B**) Temperature versus ruthenium concentration phase diagram. Blue circles – this work, black squares (*16*), red up triangles (*17*, *18*).

Initially we were motivated by the intriguing concentration independent ferromagnetic (FM) transition temperature in $CaTi_{1-x}Ru_xO_3$ (0<x<1) (*16-19*) (see Fig. 1.B). Both compounds, $CaTiO_3$ and $CaRuO_3$, crystallize in the orthorhombic modification of the ideal cubic perovskite structure. Continuous solid solution of the above compounds forms throughout full concentration range (*16, 19*), due to very small difference of effective ionic radii in octahedral coordination of $Ti^{4+}$ (0.605 Å) and $Ru^{4+}$ (0.62 Å) (*20, 21*). $CaTiO_3$ is quantum paraelectric band insulator with direct gap of 3.3 eV (*22*), where titanium ion is in 4+ valence state $Ti^{+4}$: $3d^0$ with total spin S=0. On the other hand, CaRuO$_3$ displays cooperative paramagnetic fluctuations and non-Fermi liquid metallic behavior down to the lowest temperatures (23). Ruthenium ion also adopts 4+ valence state Ru$^{4+}$: *4d$^4$ t$_{2g}^4$e$_g^0$* with total spin S=1. The understanding of FM in CaTi$_{1-x}$Ru$_x$O$_3$ has been based on Stoner criterion for FM in metals (16, 24). Namely, small variation of density of states (DOS) at the Fermi level may lead to fulfillment of the Stoner criterion and consequently to the FM ground state, according to *ID(E$_F$)>1*, where Stoner parameter *I* stands for energy gain from electron correlation, and *D(E$_F$)* is DOS at Fermi level per atom and spin orientation. However, main objections which question this scenario are: *i)* FM transition temperature does not vary with Ti impurity concentration in CaRuO$_3$, despite its sharply peaked DOS at Fermi level (24), and *ii)* The increase of concentration of Ti impurities eventually brings insulating behavior when *D(E$_F$)* reaches zero value. Experimentally, above 20% Ti per CaRuO$_3$ formula unit in thin film samples makes the system semiconducting (25).

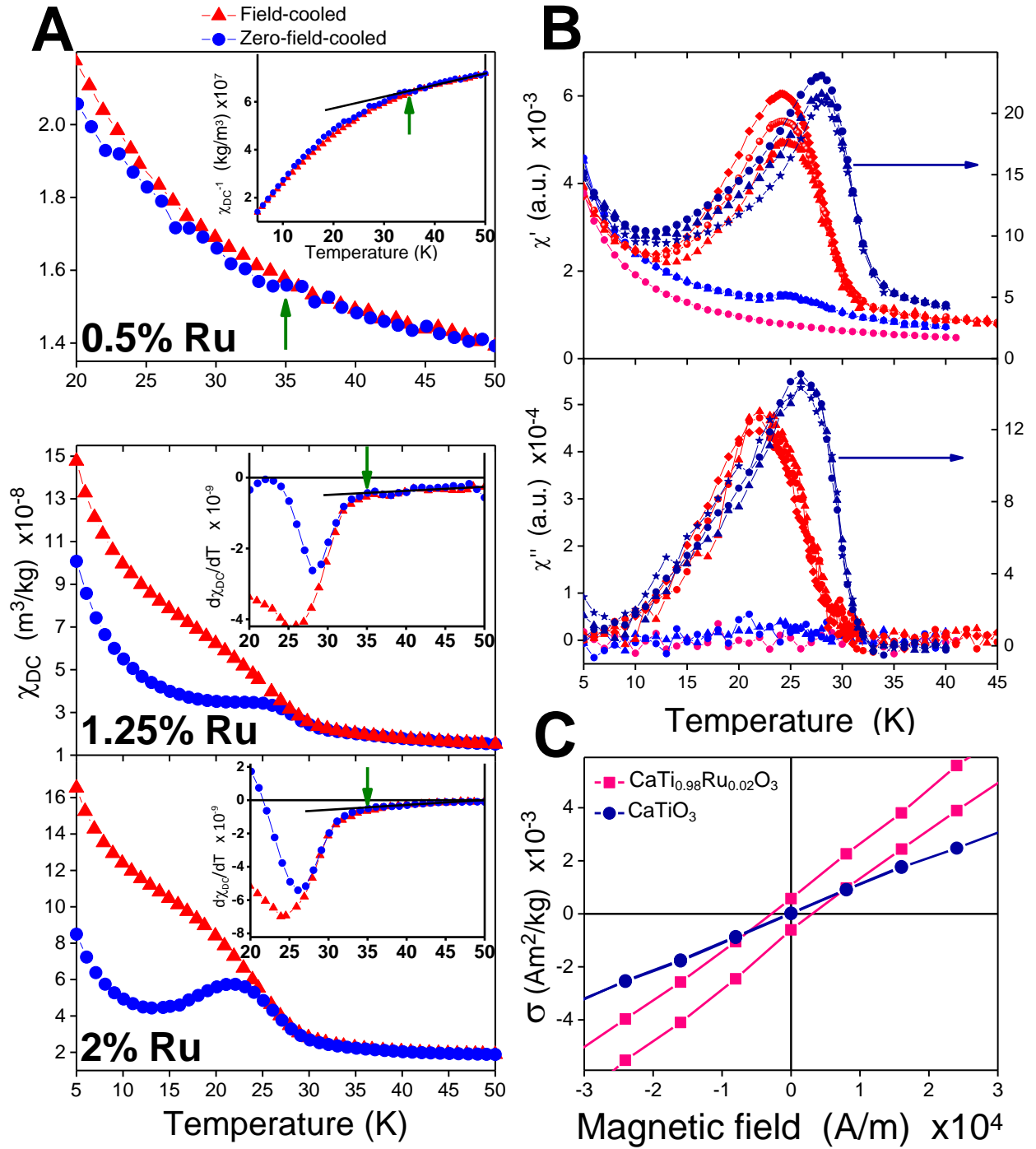

**Fig. 2. DC and AC magnetic measurements of CaTi$_{1-x}$Ru$_x$O$_3$.** (A) Up red triangles denote FC and blue circles ZFC measurements. x=0.5% - DC mass magnetization susceptibility versus T, measured in 1989 A/m applied magnetic field, and scaled to highlight divergence between ZFC and FC measurements. Inset shows inverse susceptibility versus T. x=1.25% and 2% - DC mass magnetization susceptibility versus T, measured in 1989 A/m and 3978.85 A/m applied magnetic field, respectively. Insets show d$\chi_{DC}$/dT vs. T to highlight the onset of "magnetism", marked by arrows. (B) AC magnetization measurements, upper panel - real part and lower panel - imaginary part. Colors represent samples, pink CaTiO$_3$, blue x=1.25%, red x=2%, royal blue x=12.5%. Shapes represent measuring frequencies, diamond 0.1 Hz, circle 1 Hz, up triangle 10 Hz, and star 100 Hz. The amplitude of the AC measuring field is 318.31 A/m. (C) Magnetic hysteresis measurements for CaTiO$_3$ and x=2% samples measured at T=2K. Magnetic hysteresis loop for CaTi$_{0.98}$Ru$_{0.02}$O$_3$ shows remanent mass magnetization $\sigma_R = 4.8 \times 10^{-4}$ Am$^2$/kg and coercive magnetic field $H_C = 2695$ A/m.

In order to clarify the nature of FM in this system, we have prepared ceramic samples at the Ti rich part of the phase diagram, looking for the minimum concentration of Ru impurities in CaTiO$_3$ at which FM appears. Figure 2.A shows DC magnetic susceptibility vs. temperature measurements for the lowest Ru concentrations in our experiments. The effect of substitution on magnetic response of the system is very robust, so that even 0.5% of Ru impurity per CaTiO$_3$ formula unit is enough to produce FM bellow 35 K. This manifests as divergence between zero-field-cooled (ZFC) and field-cooled (FC) branches of magnetic susceptibility, and departure from linear dependence of the inverse magnetic susceptibility, shown in the inset of Fig. 2.A for 0.5% Ru. Further increase of Ru concentration produces more pronounced FM-like phase transition, but without change in transition temperature, as marked with arrows in Fig. 2.A. In Figure 2.B we summarized the results of AC magnetic susceptibility experiments on CaTiO$_3$, 1.25%, 2%, and 12.5% Ru per formula unit. Clearly, FM behavior is observed in all Ru doped samples as compared with paramagnetic response of CaTiO$_3$. FM is also evidenced in magnetic hysteresis in 2% Ru doped sample (remanent mass magnetization $\sigma_R = 4.8 \times 10^{-4}$ *Am$^2$/kg* and coercive magnetic field $H_C = 2695$ *A/m*), shown in Fig. 2.C. Our samples apparently show parasitic paramagnetism stemming from the oxygen vacancies (26). However, apart from shielding of magnetic response, this circumstance does not influence our results.

Next, we show that for the very low doping of Ru impurities in CaTiO$_3$, no conceivable conventional magnetic interactions can explain magnetic ordering in these materials at any temperature. Assuming for simplicity, the ideal cubic perovskite crystal structure and even distribution of Ru ions within CaTiO$_3$ matrix, average distance between them is given by $<d> = a_0(1/n)^{1/3}$ (where $a_0$ is lattice constant, and *n* is Ru concentration in %). For various concentrations *n (%) = {0.5; 1.25; 2; 5}* respective inter-impurity distances are $<d>(a_0) = \{5.8; 4.3; 3.7; 2.7\}$. Evidently, magnetic dipole-dipole interaction, which decays with increasing distance as *1/r$^3$*, cannot account for the observed transition. Further, in structurally and electronically similar system SrRu$_{1-x}$Ti$_x$O$_3$ (27), photoemission spectroscopy and x-ray absorption spectroscopy experiments indicate that there are no charge fluctuations from the Ru$^{4+}$ *4d* states at Fermi energy level to the Ti$^{4+}$ *3d* states, due to large energy separation of the corresponding energy bands. This implies that any interaction between Ru ions via Ti ion is equal to zero. Accordingly, lone Ru ions imbedded in CaTiO$_3$ matrix should produce paramagnetic response. Possible clustering of magnetic ions would worsen the situation, since the distance between clusters would increase for given Ru concentration, compared to isolated single Ru ions.

Finally, in order to confirm that magnetism in our specimens is uniquely associated with quantum paraelectricity in $CaTiO_3$, we have synthesized equally doped (5% Ru per formula unit) $CaTiO_3$ and $CaZrO_3$. Unlike QPE properties of $CaTiO_3$, dielectric constant of $CaZrO_3$ has classical dependence on temperature (*28*). In Fig. 3 we compare the magnetic response of these compounds. While real and imaginary part of AC magnetic susceptibility measurements vs. temperature and mass magnetization vs. magnetic field for $CaTi_{0.95}Ru_{0.05}O_3$ (Fig. 3.A and C, respectively) clearly show FM order below $T^* \approx 35\ K$, corresponding results for $CaZr_{0.95}Ru_{0.05}O_3$ (Fig. 3. B and D) show typical paramagnetic behavior.

Because we have exhausted conventional mechanisms to explain magnetic ordering in this system, we had been forced to consider exotic multiferroic mechanism, which involves coupling between spins of magnetic impurities with QPE fluctuations. In QPE dielectric constant increases with falling temperature, following Curie-Weiss law for ferroelectrics $\varepsilon_r = C/(T-\theta)$, where $\theta$ is the Weiss temperature and $C$ is the Curie constant. At intermediate temperatures $\varepsilon_r(T)$ deviates from Curie-Weiss behavior and saturates at lower temperatures (*5*). Quantum mechanical mean field approach gives (Barrett's formula), $\varepsilon_r(T) = A + C/[(T_1/2)coth(T_1/2T)-T_0]$ (*29*). Here $A$ is a constant offset, $T_0$ represents the Curie-Weiss temperature in the classical limit, and $T_1$ is the characteristic crossover temperature between classical and quantum regime. The fitting parameters of Barrett's formula to the experimental data for $CaTiO_3$ are: $C = 7.7 \times 10^4\ K$, $T_1 = 104\ K$ and $T_0 = -159\ K$ (*26*); while dielectric constant approaches saturation level at 35 K (*26*, *30*). This is the exact temperature $T^*$, at which the onset of FM occurs in our experiments, thus underpinning our main result.

Next, we discuss a possible route towards understanding of the mechanism behind this phenomenon. A useful hint is to view polar lattice vibrations as collection of oscillating electric dipoles, which induce magnetic moment, $\boldsymbol{M} \sim \boldsymbol{r} \times \boldsymbol{j} \sim \boldsymbol{P} \times \partial_t \boldsymbol{P}$. This approach has been recently used to describe magnetization induced by time dependent polarization, and termed "dynamical multiferroicity" (*31*). For example, according to this study, magnetization arising from two perpendicular degenerate optical phonons is static and takes the form $\boldsymbol{M}(0) \sim \omega_0 sin(\phi) A_1 A_2 \boldsymbol{z}$. It is oriented normal to both phonon modes, and depends only on frequency $\omega_0$, phase shift between phonons $\phi$, and their amplitudes $A_1$ and $A_2$. In general case, when phonons have different frequencies, the induced magnetization will oscillate in time. Thus induced magnetization, generated by time varying polarization, then couples with impurity magnetic moment, leading to the observed magnetism in our experiments.

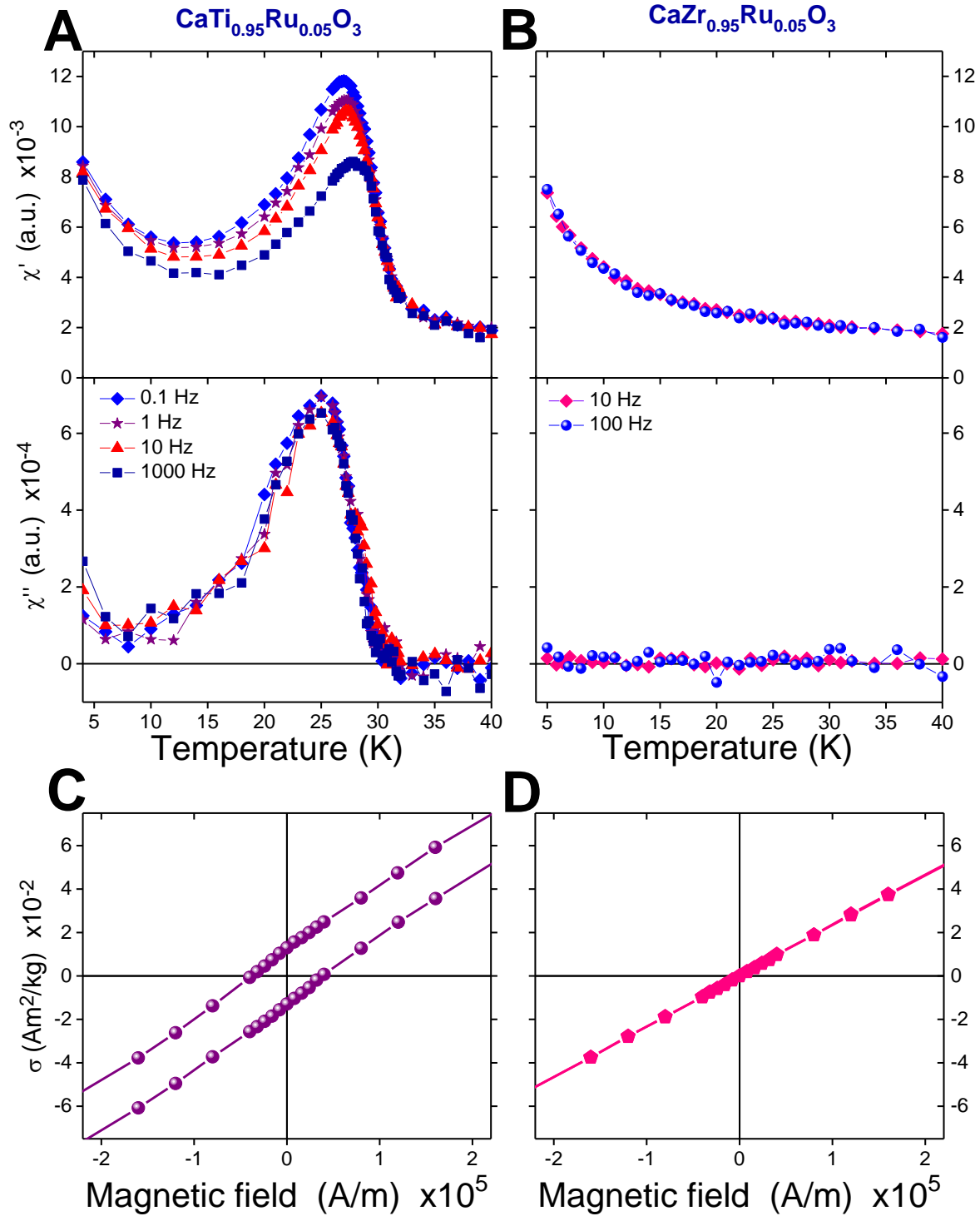

**Fig. 3. Comparison of magnetic properties of $CaTi_{0.95}Ru_{0.05}O_3$ and $CaZr_{0.95}Ru_{0.05}O_3$.** Temperature dependence of the real (upper panel) and imaginary (lower panel) AC magnetic susceptibility of **(A)** $CaTi_{0.95}Ru_{0.05}O_3$ and **(B)** $CaZr_{0.95}Ru_{0.05}O_3$. The amplitude of the AC measuring field is 318.31 A/m. Mass magnetization versus magnetic field at T=2 K of **(C)** $CaTi_{0.95}Ru_{0.05}O_3$ and **(D)** $CaZr_{0.95}Ru_{0.05}O_3$. Magnetic hysteresis loop for $CaTi_{0.95}Ru_{0.05}O_3$ shows remanent mass magnetization $\sigma_R = 1.3 \times 10^{-2}$ $Am^2/kg$ and coercive magnetic field $H_C = 0.4 \times 10^5$ A/m.

Now, we can view the Ru impurity as a probe to detect intrinsic properties of pure CaTiO$_3$. From this stance, as magnetism occurs only when the system enters quantum regime, it follows that CaTiO$_3$ undergoes finite temperature phase transition at $T^*$ from paraelectric phase for $T > T^*$ to QPE phase for $T < T^*$ (Fig. 1B). Because quantum phase transitions do not involve entropy, QPE phase has also to be ordered, thus representing quantum coherent state, similar to superfluid phase of helium-4 and superconductivity. This concept has been proposed to explain anomaly at 37 K in electron paramagnetic resonance of Fe$^{3+}$ in SrTiO$_3$ (*32*). Subsequently, the existence of this anomaly has been confirmed by various experimental techniques (see for example (*33*)), but also concerns about the nature of the anomaly have been raised, most notably expressed in Ref. 34. Namely, resonant piezoelectric spectroscopy measurements showed strong polar resonances below 40 K. These resonances have been attributed to ferroelectric polarization, located in the ferroelastic domain walls. Our ceramic samples undoubtedly contain oxygen vacancies, which tend to be trapped in ferrielectric twin walls (*35*). Furthermore, grain boundaries may also cause local electric polarization. However, these effects cannot contribute to the observed magnetism, since, as explained above, only time varying electric polarization can induce dynamic multiferroicity.

We expect that spin-QPE-fluctuations-coupling effect, or more generally spin-charge-dynamics-coupling, will turn out to be ubiquitous and account for the previously unrecognized magnetism in many materials, especially in strongly anisotropic systems like quasi-two-dimensional organics (*15*), interfaces and surfaces. This leads to perspective of designing materials with tailored properties for use in novel technologies. Finally, our results provide a powerful tool for studies of exotic phases in quantum ferroelectrics.

**Acknowledgments:** I. M. Bradarić is grateful for financial support from Ministry of education, science, and technological development, Republic of Serbia (Grant No. 171027).